\documentclass[reprint,aps,prd,twocolumn,notitlepage,showpacs,nofootinbib,preprintnumbers,superscriptaddress]{revtex4-1}
\usepackage{amsmath}
\usepackage{amsfonts}
\usepackage{amssymb}
\usepackage{latexsym}
\usepackage{enumerate}
\usepackage{color,xcolor}
\usepackage{graphicx}
\usepackage{bm}
\usepackage{epsfig}
\usepackage[english]{babel}
\usepackage{hyperref}
\usepackage{times}
\usepackage{comment}
\usepackage{epstopdf}

\def\d{\mathrm{d}}
\def\bol{\mathrm{bol}}
\def\eff{\mathrm{eff}}
\def\isco{\mathrm{ISCO}}
\def\dom{\left(\d\theta^2+\sin^2\theta \d\varphi^2\right)}
\def\rh{r_\mathrm{H}}
\def\rext{r_\mathrm{ext}}
\def\risco{r_\mathrm{ISCO}}
\def\rin{r_\mathrm{in}}
\def\rout{r_\mathrm{out}}

\begin{document}
%\preprint{\vbox{ \hbox{}   \hbox{} }}

\title{Accretion disks around magnetically charged black holes in string theory with an Euler-Heisenberg correction}
\author{Yu-Hao Jiang}
\author{Towe Wang}
\email[Electronic address: ]{twang@phy.ecnu.edu.cn}
\affiliation{Department of Physics, East China Normal University, Shanghai 200241, China\\}
\date{\today\\ \vspace{1cm}}
\begin{abstract}
A relativistic model of geometrically thin and optically thick accretion disks is worked out for a class of magnetically charged black holes, which are solutions to the Einstein-Maxwell-dilaton theory with a dilaton-coupled Euler-Heisenberg correction. Constraints on black hole parameters are derived from the subextremality condition and energy conditions. The Keplerian orbits, the time-averaged energy flux, the local temperature, the specific luminosity and the conversion efficiency of accreting mass into radiation are obtained and compared with Schwarzschild and Gibbons-Maeda-Garfinkle-Horowitz-Strominger black holes.
\end{abstract}

%\pacs{}

\maketitle

%\tighten

%%%%%%%%%%%%%%%%%%%%%%%%%%%%%%%%%%%%%%%%%%

%\tableofcontents

\section{Introduction}\label{sect-intro}
Breakthroughs in the last decade \cite{LIGOScientific:2016vlm,GRAVITY:2018ofz,EventHorizonTelescope:2019dse} have opened the era of precision black-hole observations. It will be promising to seek in high-precision astronomical observations for clues of untested physical theories, such as string theory, Lovelock theory and nonlinear electrodynamics. The Gibbons-Maeda-Garfinkle-Horowitz-Strominger (GMGHS) black hole \cite{Gibbons:1982ih,Gibbons:1987ps,Garfinkle:1990qj} is a simple and typical black hole solution in low-energy string theory. It is an exact solution to the Einstein-Maxwell-dilaton theory, which can be regarded as a truncated four-dimensional low-energy limit of the heterotic string theory \cite{Garfinkle:1990qj,Gross:1985rr,Sen:1992ua}.

String theory was originally developed as a phenomenological model of hadronic interactions, but now is the most promising candidate theory of quantum gravity. In Ref. \cite{Gross:1985rr}, it was shown that the ten-dimensional $N=1$ supergravity coupled to a Yang-Mills field can be derived as a low-energy limit from the heterotic string theory. As mentioned in Ref. \cite{Sen:1992ua}, the bosonic sector of this low-energy effective theory can be further truncated to a four-dimensional Einstein-Maxwell-dilaton theory by compactifying six of the ten dimensions and including only a $U(1)$ component of the Yang-Mills field. Therefore, as an exact solution to the Einstein-Maxwell-dilaton theory, the GMGHS black hole provides us with an opportunity to look for astrophysical hints of string theory.

Over nearly four decades, modifications or corrections to the GMGHS solution have been studied exactly \cite{Gibbons:1987ps,Sen:1992ua,Chan:1996nk,Cai:1997ii,Cai:2004iy,Gao:2004tu,Yazadjiev:2005du}, perturbatively \cite{Natsuume:1994hd,Cano:2021nzo,Zatti:2023oiq}, and numerically \cite{Ohta:2012ih} for various extensions of the Einstein-Maxwell-dilaton theory. In these solutions, an essential role is played by the magnetic charge, without which the GMGHS solution will reduce to the ordinary Schwarzschild solution. The astrophysical community generally ignores the electrically charged black holes, because they should be quickly neutralized by accreting ionized plasma. Unlike the electrically charged counterpart, magnetically charged black holes can survive and be relevant to astrophysics, owing to the paucity of magnetically charged matter \cite{Bai:2020spd,Ghosh:2020tdu,Diamond:2021scl}. In fact, magnetically charged black holes are of interest not only to string theory, but also to the standard electroweak theory \cite{Lee:1994sk,Maldacena:2020skw}.

Very recently, inspired by string theory and Lovelock theory, the authors of Ref. \cite{Bakopoulos:2024hah} proposed an extension of the Einstein-Maxwell-dilaton theory by coupling dilaton with Euler-Heisenberg electrodynamics \cite{Heisenberg:1936nmg} in a specified way. In their proposed model, they found a new black hole solution which resembles a magnetically charged black hole in Einstein-Euler-Heisenberg theory \cite{Yajima:2000kw,Ruffini:2013hia} and can recover the GMGHS black hole in a certain limit. Although the Euler-Heisenberg electrodynamics is a classical theory of nonlinear electrodynamics, it can be also derived as an effective theory from one-loop corrections in quantum electrodynamics (QED) \cite{Ruffini:2013hia}. In some sense, both stringy and QED effects are built in this model and in the black hole solution, so the model and the black hole deserve further investigations. Timelike geodesics, energy conditions, thermodynamics and linear perturbations of the new black hole were discussed in Ref. \cite{Bakopoulos:2024hah}. Null geodesics, strong gravitational lensing effects of this black hole and constraints on model parameters from the Event Horizon Telescope \cite{EventHorizonTelescope:2019dse} were investigated in Ref. \cite{Vachher:2024fxs}. The new solution allows for black holes with the same horizon radius but different charges \cite{Bakopoulos:2024hah}, whose shadow images were compared in Ref. \cite{Xu:2024gjs}.

In this paper, our focus is on the model of accretion disk around the new black hole. In realistic environments, accretion disks can play crucial roles in black hole astrophysics \cite{Shakura:1972te,Novikov:1973kta,Page:1974he,Thorne:1974ve,Yuan:2014gma}. For example, it was widely held that the shadow size in a black hole's image is completely determined by mass and spin parameters of the black hole \cite{Synge:1966okc,Bardeen:1973,Luminet:1979nyg}, but recent investigations show that this conclusion relies on accretion models, especially on the radius of its inner edge and the motion of the accreting plasma \cite{Gralla:2019xty,Narayan:2019imo,Qu:2023hsy}. In contrast, both refractive and pure gravitational effects of the accreting plasma to the shadow size are usually negligible for near-future observations; see Ref. \cite{Perlick:2023znh} and references therein.

In the past few years, accretion models of the GMGHS black hole and modified spacetimes have attracted much attention. In Ref. \cite{Karimov:2018whx}, the authors studied the emissivity properties of thin accretion disks around GMGHS black holes, compared their behaviors in the Einstein and string frames with those of Reissner-Nordstr\"{o}m and Schwarzschild black holes, and noticed the differences in the extreme limit that may lead to naked singularities and wormholes. In Ref. \cite{Heydari-Fard:2020ugv}, authors investigated the effects of dilaton and rotation parameters on the properties of thin accretion disks around static and slowly rotating charged dilaton black holes under Einstein-Maxwell-dilaton gravity, and compared the results with Schwarzschild and Kerr black holes. In Ref. \cite{Banerjee:2020ubc}, authors explored the properties of jet power and radiative efficiency in microquasars within the framework of Einstein-Maxwell-dilaton-axion (EMDA) gravity, and revealed a preference for Kerr black holes over Kerr-Sen black holes with dilaton charges based on observational data using chi-square analysis. In Ref. \cite{Banerjee:2020qmi}, authors found that the dilaton parameter $r_2\sim0.2$ is preferred by comparing the theoretical luminosity from the accretion disk in the Kerr-Sen background with the optical observations of quasars. In Ref. \cite{Feng:2023iha}, authors explored the choked accretion of ultrarelativistic fluids onto Kerr-Sen black holes in EMDA gravity, derived solutions for the velocity potential and examined the influence of various parameters on accretion and ejection rates, radiative efficiency and redshift. In Ref. \cite{Alloqulov:2024zln}, authors studied the influence of nonrotating black hole parameters on the basic properties of thin accretion disks under the Einstein-Maxwell-scalar theory, which provides a reference for future astronomical observations. In Ref. \cite{Feng:2024iqj}, authors studied the accretion process of a supermassive black hole in EMDA gravity, and used observational data from M87* and Sgr A* to constrain parameters of the model.
%In Ref. \cite{Claros:2024atw}, authors presented new analytical formulas for accurately describing light rays in spherically symmetric static spacetimes, and applied these formulas to analyze images and polarization of black hole accretion disks.

Early this year, the Novikov-Thorne model of accretion \cite{Novikov:1973kta,Page:1974he,Thorne:1974ve} was applied to the Einstein-Euler-Heisenberg black hole. In Refs. \cite{Ma:2024oqe,You:2024uql}, the authors studied the exact solution of Einstein-Euler-Heisenberg black hole surrounded by a perfect fluid dark matter or a cold dark matter halo as well as the properties of the thin accretion disk, which provides a new perspective for us to study dark matter.

%neglect the scattering and absorption of photons by plasma and

In this paper, following the procedure of Refs. \cite{Karimov:2018whx,Alloqulov:2024zln,Feng:2024iqj}, we will construct a simple Novikov-Thorne model of thin accretion disks around the magnetically charged black holes in string theory with the Euler-Heisenberg correction \cite{Bakopoulos:2024hah}. The paper is organized in the following way. In Sec. \ref{sect-bh}, we review and clarify the string-inspired model proposed in Ref. \cite{Bakopoulos:2024hah} and the magnetically charged black hole in this model. Constraints on black hole parameters from energy conditions are worked out in Sec. \ref{sect-ec}. Then the thin accretion disk of this black hole is investigated, with Sec. \ref{sect-orb} being devoted to its geometrical properties and Sec. \ref{sect-rad} to its radiative properties. Finally, in Sec. \ref{sect-out}, we summarize the obtained results and give a future outlook related to the present work.

We will work in the Einstein frame and the geometrized unit system $G=c=\hbar=k=1$ unless otherwise stated. Here $G$, $k$ and $\hbar$ are the gravitational, Boltzmann and reduced Planck constants, and $c$ is the speed of light. Throughout this paper, the prime denotes differentiation with respect to an areal radial coordinate $r$. In Secs. \ref{sect-orb} and \ref{sect-rad}, diagonal metrics are assumed, $g_{\varphi\varphi}$ is evaluated at $\theta=\pi/2$, and thus the spherical radial coordinate can be understood as the planar radial coordinate. In all numerical simulations, we will set the black hole mass to unity $M=1$.

\section{Magnetically charged black holes in string theory with an Euler-Heisenberg correction}\label{sect-bh}
As one of the low-energy limits of string theory, the Einstein-Maxwell-dilaton theory has the following form of action in the Einstein frame \cite{Garfinkle:1990qj,Gross:1985rr}:
\begin{equation}\label{actEMD}
\mathcal{S}_{\mathrm{EMD}}=\frac{1}{16\pi}\int\d x^4\sqrt{-g}\left[\mathcal{R}-2\nabla^\mu\phi\nabla_\mu\phi-e^{-2\phi}F_{\mu\nu}F^{\mu\nu}\right],
\end{equation}
in which $\mathcal{R}$ is the Ricci scalar, $\phi$ is a dilaton field, $\nabla$ denotes the covariant derivative, the indices $\mu,\nu=0,1,2,3$, and $F_{\mu\nu}=\partial_{\mu}A_{\nu}-\partial_{\nu}A_{\mu}$ is the gauge field strength.

Inspired by string theory and Lovelock theory, Bakopoulos \emph{et. al.} proposed to augment action \eqref{actEMD} with a dilaton-coupled nonlinear Euler-Heisenberg term \cite{Bakopoulos:2024hah}
\begin{eqnarray}\label{actfEH}
\nonumber\mathcal{S}_{f\mathrm{EH}}&=&-\frac{1}{16\pi}\int\d x^4\sqrt{-g}f(\phi)\Bigl[2\alpha F^{\mu}_{~\nu}F^{\nu}_{~\rho}F^{\rho}_{~\sigma}F^{\sigma}_{~\mu}\\
&&-\beta\left(F_{\mu\nu}F^{\mu\nu}\right)^2\biggr],
\end{eqnarray}
and they specified the coupling function to
\begin{equation}\label{coupf}
f(\phi)=-\frac{1}{2}\left(3e^{-2\phi}+3e^{2\phi}+4\right).
\end{equation}
The full action of their model is $\mathcal{S}=\mathcal{S}_{\mathrm{EMD}}+\mathcal{S}_{f\mathrm{EH}}$. In this model, they successfully get a new solution of black hole characterized by the mass $M$, the magnetic charge $Q$ and the difference between model parameters $\alpha-\beta$.
%It resembles Einstein-Euler-Heisenberg-dilaton theory.

Before presenting their solution, we would like to rewrite the coupling term \eqref{actfEH} in a more enlightening form \cite{Yajima:2000kw}. According to Eq. (16) in Ref. \cite{Escobar:2013rsa}, there is an identity
\begin{equation}
F^{\mu}_{~\nu}F^{\nu}_{~\rho}F^{\rho}_{~\sigma}F^{\sigma}_{~\mu}=\frac{1}{2}\left(F_{\mu\nu}F^{\mu\nu}\right)^2+\frac{1}{4}\left(F_{\mu\nu}\tilde{F}^{\mu\nu}\right)^2,
\end{equation}
where $\tilde{F}_{\mu\nu}=\epsilon_{\mu\nu\rho\sigma}F^{\mu\nu}/2$ is the Hodge dual of field strength. Plugging this identity to Eq. \eqref{actfEH}, we find
\begin{eqnarray}\label{actEH}
\nonumber\mathcal{S}_{f\mathrm{EH}}&=&-\frac{1}{16\pi}\int\d x^4\sqrt{-g}f(\phi)\biggl[(\alpha-\beta)\left(F_{\mu\nu}F^{\mu\nu}\right)^2\\
&&+\frac{\alpha}{2}\left(F_{\mu\nu}\tilde{F}^{\mu\nu}\right)^2\biggr].
\end{eqnarray}
It is a linear combination of squares of two independent Lorentz invariants, the scalar $F_{\mu\nu}F^{\mu\nu}$ and the pseudoscalar $F_{\mu\nu}\tilde{F}^{\mu\nu}$. Then one can recover the Einstein-Euler-Heisenberg action in Ref. \cite{Yajima:2000kw} by taking $\alpha-\beta=a/5$, $\phi=0$ in Eqs. \eqref{actEMD} and \eqref{actEH}. Alternatively, if one sets $\alpha=\beta=0$, the model will reduce to the Einstein-Maxwell-dilaton theory \cite{Garfinkle:1990qj}. In Ref. \cite{Yajima:2000kw}, the coefficient of the $(F_{\mu\nu}\tilde{F}^{\mu\nu})^2$ term in the Einstein-Euler-Heisenberg action does not appear in its magnetically charged black hole solution. In view of this, it is reasonable to expect the magnetically charged black hole solution to model $\mathcal{S}=\mathcal{S}_{\mathrm{EMD}}+\mathcal{S}_{f\mathrm{EH}}$ will depend solely on $\alpha-\beta$. More generally, on a time slice with an electric field $\mathbf{E}$ and a magnetic field $\mathbf{B}$, we have $F_{\mu\nu}F^{\mu\nu}=2(\mathbf{B}\cdot\mathbf{B}-\mathbf{E}\cdot\mathbf{E})$ and $F_{\mu\nu}\tilde{F}^{\mu\nu}=-4\mathbf{E}\cdot\mathbf{B}$, and the latter vanishes as $\mathbf{E}=0$.

In Ref. \cite{Bakopoulos:2024hah}, the new solution of the magnetically charged black hole was presented in two coordinate systems. For our purpose in this paper, we will only recapitulate it in the physical coordinate system, in which it takes the form \cite{Bakopoulos:2024hah}
\begin{equation}\label{metric}
\d s^2=-B(r)\d t^2+\frac{W(r)^2}{B(r)}\d r^2+r^2\dom,
\end{equation}
and the functions $B(r)$, $W(r)$ and the fields $\phi(r)$, $A_{\mu}$ are given by
\begin{eqnarray}
\label{B}B(r)&=&1-\frac{4M^2}{\sqrt{Q^4+4M^2r^2}+Q^2}-\frac{2(\alpha-\beta)Q^4}{r^6},\\
\label{W}W(r)^2&=&\frac{4M^2r^2}{Q^4+4M^2r^2},\\
\label{phi}\phi(r)&=&-\frac{1}{2}\ln\frac{\sqrt{Q^4+4M^2r^2}-Q^2}{\sqrt{Q^4+4M^2r^2}+Q^2},\\
\label{A}A_{\mu}&=&\left(0,0,0,Q\cos\theta\right).
\end{eqnarray}
The radial coordinate in the line element \eqref{metric} is also known as the ``areal radius,'' because a sphere of constant $r$ has an area of $4\pi r^2$. This coordinate system will be more convenient for us to study the radiative flux and the luminosity later. It is noteworthy that, in the limit $\alpha-\beta=0$, the solution \eqref{metric} reduces to the GMGHS black hole. In general, the black holes have one horizon for $\alpha-\beta\geq0$ and two horizons for $\alpha-\beta<0$ \cite{Bakopoulos:2024hah,Vachher:2024fxs}.

\begin{figure}[htbp]
    \centering
    \includegraphics[width=0.45\textwidth]{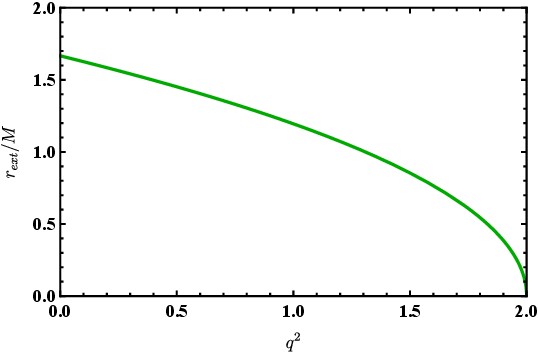}\\
    \includegraphics[width=0.45\textwidth]{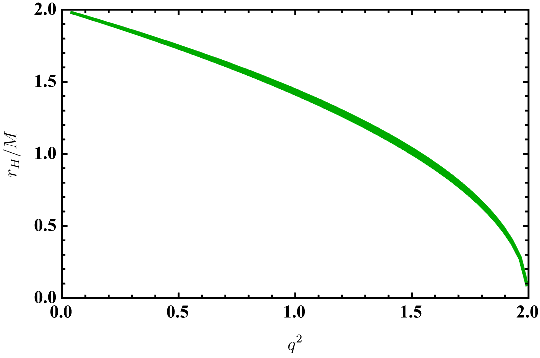}
    \caption{Extremal radius of degenerate horizons (green curve in the top panel, $\alpha-\beta<0$) and radius of the single horizon respecting energy conditions (green region in the bottom panel, $\alpha-\beta>0$) against $q^2$.
    }\label{fig:rh}
\end{figure}

\begin{figure*}[htbp]
    \centering
    \includegraphics[width=0.45\textwidth]{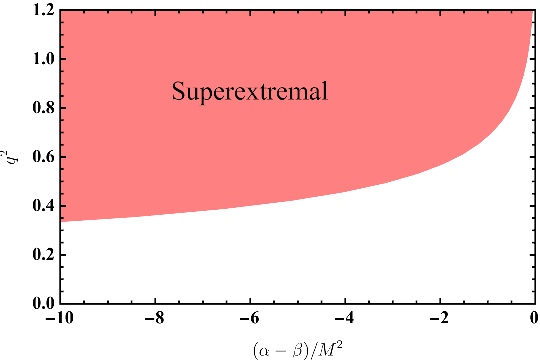}\includegraphics[width=0.45\textwidth]{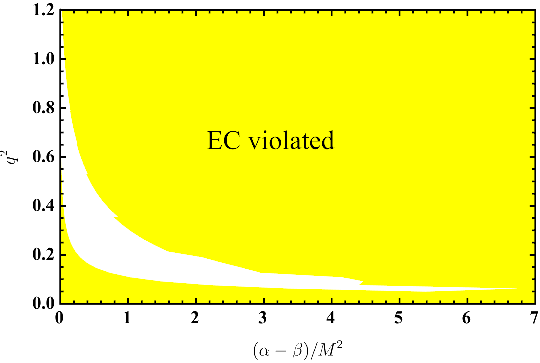}\\
    \caption{Parameter regions violating the subextremal condition (shaded red in the left panel) and energy conditions (shaded yellow in the right panel) discussed in Secs. \ref{sect-bh} and \ref{sect-ec}. Conversely, blank areas respect both the subextemality condition (if applicable) and energy conditions. Notice the left panel is in accordance with Ref. \cite{Vachher:2024fxs}. %Critical values of black hole charge (top panel) and extreme radius of the horizon (bottom panel) as functions of the parameter difference $\alpha-\beta$.
    }\label{fig:ext}
\end{figure*}

In the latter case, if the two horizons are degenerate, then the solution describes extremal black holes, and we have $B=B'=0$, from which one can infer the radius of degenerate horizons
\begin{eqnarray}\label{rext}
\nonumber\frac{\rext^2}{M^2}&=&\sqrt{\left(\frac{1}{8}q^4+\frac{2}{3}q^2-\frac{25}{18}\right)^2-\frac{1}{3}q^4\left(q^2-2\right)}\\
&&-\left(\frac{1}{8}q^4+\frac{2}{3}q^2-\frac{25}{18}\right)
\end{eqnarray}
with $q=Q/M$ as well as the subextremality condition
\begin{equation}\label{Qext}
\frac{\alpha-\beta}{M^2}<-\frac{M^{-4}\rext^4\left(\sqrt{q^4+4M^{-2}\rext^2}-q^2\right)^2}{12q^4\sqrt{q^4+4M^{-2}\rext^2}}.
\end{equation}
According to Eq. \eqref{rext}, the radius of extremal black holes is plotted as a function of $q^2$ in the top panel of Fig. \ref{fig:rh}. It decreases from $5M/3$ to zero as the absolute value of $Q$ grows from zero to $\sqrt{2}M$. The subextremality condition is illustrated in the left panel of Fig. \ref{fig:ext} and will be considered in our simulations later. As $(\alpha-\beta)/M^2$ approaches zero from the negative direction, the extremal value of $|Q|$ goes to $\sqrt{2}M$ from zero.


%Ref. \cite{Bakopoulos:2024hah} has studied the null, weak and strong energy conditions outside the event horizon of the black hole \eqref{metric}. Making use of their analytical results, we prove in Appendix \ref{sect-ec} that the necessary and sufficient condition for these energy conditions is \eqref{SEC} with $r$ evaluated on the event horizon.

\section{Energy conditions outside event horizons}\label{sect-ec}
In the physical coordinate system, Ref. \cite{Bakopoulos:2024hah} has studied energy conditions outside the event horizon of the black hole \eqref{metric} in string theory with an Euler-Heisenberg correction. It was found that the energy density, the radial pressure and the tangential pressure for the stress tensor of their model are
\begin{eqnarray}
\label{rho}\rho&=&\frac{B}{W^2}\phi'^2+\frac{Q^2}{r^4}e^{-2\phi}+\frac{2(\alpha-\beta)Q^4}{r^8}f(\phi),\\
\label{pr}p_r&=&\frac{B}{W^2}\phi'^2-\frac{Q^2}{r^4}e^{-2\phi}-\frac{2(\alpha-\beta)Q^4}{r^8}f(\phi),\\
\label{pt}p_{\theta}&=&-\frac{B}{W^2}\phi'^2+\frac{Q^2}{r^4}e^{-2\phi}+\frac{6(\alpha-\beta)Q^4}{r^8}f(\phi),
\end{eqnarray}
respectively. Imposing the null, weak and strong energy conditions on the density and pressures together, one has \cite{Bakopoulos:2024hah}
\begin{eqnarray}
\label{EC1}\rho+p_r&\geq&0,\\
\label{EC2}\rho+p_{\theta}&\geq&0,\\
\label{EC3}\rho&\geq&0,\\
\label{EC4}\rho+p_r+2p_{\theta}&\geq&0.
\end{eqnarray}

Outside event horizons, $B>0$ and $f(\phi)<0$; hence, both the first and second terms in Eq. \eqref{rho} are positive, while the third term is negative for $\alpha-\beta>0$ but positive for $\alpha-\beta<0$. Based on this observation, we can infer that all of the above energy conditions are satisfied when $\alpha-\beta\leq0$. However, in the case $\alpha-\beta>0$, the conditions \eqref{EC2}, \eqref{EC3} and \eqref{EC4} are nontrivial, taking the form
\begin{eqnarray}
\label{EC5}\frac{Q^2}{r^4}e^{-2\phi}+\frac{4(\alpha-\beta)Q^4}{r^8}f(\phi)&\geq&0,\\
\label{EC6}\frac{B}{W^2}\phi'^2+\frac{Q^2}{r^4}e^{-2\phi}+\frac{2(\alpha-\beta)Q^4}{r^8}f(\phi)&\geq&0,\\
\label{EC7}\frac{Q^2}{r^4}e^{-2\phi}+\frac{6(\alpha-\beta)Q^4}{r^8}f(\phi)&\geq&0.
\end{eqnarray}
Among them the tightest constraint is \eqref{EC7}. After substitutions of the coupling function \eqref{coupf} and the dilaton field \eqref{phi}, it can be rewritten as
%\begin{eqnarray}
%\nonumber&&\frac{r^4}{Q^2}e^{-2\phi}-3(\alpha-\beta)\left(3e^{-2\phi}+3e^{2\phi}+4\right)\geq0\\
%\nonumber&&\frac{r^4}{Q^2}\frac{\left(\sqrt{Q^4+4M^2r^2}-Q^2\right)^2}{4M^2r^2}\geq3(\alpha-\beta)\left[\frac{6\left(2Q^4+4M^2r^2\right)}{4M^2r^2}+4\right]\\
%\nonumber&&\frac{r^2}{4M^2Q^2}\left(\sqrt{Q^4+4M^2r^2}-Q^2\right)^2\geq3(\alpha-\beta)\left(\frac{3Q^4}{M^2r^2}+10\right).
%\end{eqnarray}
\begin{equation}\label{SEC}
\alpha-\beta\leq\frac{\left(\sqrt{Q^4+4M^2r^2}-Q^2\right)^2}{\frac{12M^2Q^2}{r^2}\left(\frac{3Q^4}{M^2r^2}+10\right)}.
\end{equation}

It is crucial to observe that the right-hand side of \eqref{SEC} is a monotonically increasing function of $r^2$, so the energy conditions can be met in the causal region outside the event horizon if and only if the condition \eqref{SEC} is satisfied on the event horizon, $r=\rh$. Obviously, this condition is well satisfied when $\alpha-\beta\leq0$. Therefore, after replacement $r\rightarrow\rh$, inequality \eqref{SEC} becomes a necessary and sufficient condition for the null, weak and strong energy conditions to hold outside event horizons.

It is interesting but laborious to examine how the model parameters $\alpha-\beta$ and the black hole parameters $M$, $Q$ are constrained by this condition, especially in the case $\alpha-\beta>0$. In a rectangular region spanned by $(\alpha-\beta)/M^2$ and $Q^2/M^2$, this can be done numerically by dividing the parameter region into tiny squares and checking \eqref{SEC} with the positive root of $B(r)=0$ in each square. Notice that $B(r)$ is a monotonically increasing function of $r^2$ and has only one positive root when $\alpha-\beta>0$.

To save labor and to get more insights, we introduce a new variable
\begin{equation}
\Delta=\frac{\sqrt{Q^4+4M^2\rh^2}-Q^2}{2M^2}
\end{equation}
to trade off $\rh^2/M^2=\Delta\left(\Delta+q^2\right)$ and rewrite $B(\rh)=0$ and \eqref{SEC} on the event horizon as
%\begin{equation}
%\alpha-\beta=\frac{r^4}{2Q^4}\left[r^2-\left(\sqrt{Q^4+4M^2r^2}-Q^2\right)\right].
%\end{equation}
\begin{eqnarray}
&&\frac{\alpha-\beta}{M^2}=\frac{\Delta^2\left(\Delta+q^2\right)^2}{2q^4}\left[\Delta^2+(q^2-2)\Delta\right],\\
&&\frac{\alpha-\beta}{M^2}\leq\frac{\Delta^2}{\frac{3q^2}{\Delta\left(\Delta+q^2\right)}\left[\frac{3q^4}{\Delta\left(\Delta+q^2\right)}+10\right]}.
\end{eqnarray}
Since $\Delta>0$, they can be combined as a cubic inequality of $\Delta$,
%\begin{equation}
%\left[\frac{9}{2}q^2+\frac{15}{q^2}\Delta\left(\Delta+q^2\right)\right]\left(\Delta+q^2-2\right)\leq\Delta.
%\end{equation}
\begin{equation}\label{rh}
\Delta^3+2(q^2-1)\Delta^2+q^2\left(\frac{13}{10}q^2-\frac{31}{15}\right)\Delta+\frac{3}{10}q^4(q^2-2)\leq0.
\end{equation}
Taking $\alpha-\beta>0$, i.e. $\Delta+q^2-2>0$ also into consideration, we present the allowed region for $\rh/M$ versus $q^2$ in the bottom panel of Fig. \ref{fig:rh}, and for $q^2$ versus $(\alpha-\beta)/M^2$ in the right panel of Fig. \ref{fig:ext}. Both analytical calculations and numerical simulations indicate that, as $(\alpha-\beta)/M^2$ approaches zero from the positive direction, the value of $|Q|$ approaches $\sqrt{2}M$ and $\rh$ tends to zero. In the other limit $Q\rightarrow0$, we find the horizon radius $\rh$ goes to $2M$ and the value of $(\alpha-\beta)/M^2$ is divergent as $8/(15q^2)$.

It deserves emphasis that $\alpha$ and $\beta$ are not charges of black holes but theoretical parameters of the model \cite{Bakopoulos:2024hah}. In principle their values can be calculated from string theory and Lovelock theory, but that is a very challenging task. In Ref. \cite{Adams:2006sv} it was claimed that $\alpha>0$ and $\alpha-\beta>0$ in any consistent effective theory. Although the energy conditions investigated above are useful assumptions in mathematical relativity, they are questionable physically as elaborated in Refs. \cite{Barcelo:2002bv,Kontou:2020bta}. In the following simulations, we will adopt an empirical attitude and treat the parameters phenomenologically in line with Ref. \cite{Bakopoulos:2024hah}. Specifically, in numerical simulations we will set $M=1$ and study five typical cases:
\begin{enumerate}[(i)]
\item $Q=0$ (red solid curves);
\item $Q=0.5$, $\alpha-\beta=0$ (blue dashed curves);
\item $Q=0.5$, $\alpha-\beta=-20$ (black solid curves);
\item $Q=0.5$, $\alpha-\beta=1$ (purple dash-dotted curves);
\item $Q=0.5$, $\alpha-\beta=200$ (orange or red dotted curves).
\end{enumerate}
Obviously, case (i) recovers the Schwarzschild black hole, while case (ii) is a GMGHS black hole with $Q=M/2$. Case (iii) represents a subextremal black hole, whereas cases (iv) and (v) are examples of single-horizon black holes meeting and violating energy conditions, respectively. Since the powers of $Q$ are even in line element \eqref{metric} and we will consider neither test charges nor external fields, all results in this paper are unchanged by reversing the sign of $Q$, for example, by choosing $Q=-0.5$ instead.

\section{Circular orbits of massive particles}\label{sect-orb}
In the model of geometrically thin accretion disks around black holes \cite{Shakura:1972te,Novikov:1973kta,Page:1974he,Thorne:1974ve}, it is assumed that the accreting materials are distributed near an axially symmetric plane, or quantitatively, inside $\cot\theta\ll1$ in the physical coordinate system \eqref{metric}. Furthermore, it is usually assumed that the disk materials are moving in nearly geodesic circular orbits in the equatorial plane $\theta=\pi/2$, and the inner edge of the disk is located at the innermost stable circular orbit (ISCO). For this reason, let us examine the ISCO and general circular obits of massive particles around the black hole \eqref{metric} in the current section and turn to radiative properties of the disk in the next section.

\subsection{Innermost stable circular orbit}\label{subsect-isco}
To determine the ISCO radius, we should study the equations of motion of a massive particle in the equatorial plane. This has been done partly in Ref. \cite{Bakopoulos:2024hah} in a different coordinate system. Here we will complete the analysis in the physical coordinate system, which will be more complicated algebraically but clearer physically.

For orbits in the equatorial plane $\theta=\pi/2$, the effective Hamiltonian of a massive particle is equal to
\begin{eqnarray}\label{ham}
\nonumber\mathcal{H}_{\eff}&=&\frac{1}{2}\left[g_{tt}\left(\frac{\d t}{\d\tau}\right)^2+g_{rr}\left(\frac{\d r}{\d\tau}\right)^2+g_{\varphi\varphi}\left(\frac{\d\varphi}{\d\tau}\right)^2\right]\\
\nonumber&=&\frac{1}{2}\left[-B(r)\left(\frac{\d t}{\d\tau}\right)^2+\frac{W(r)^2}{B(r)}\left(\frac{\d r}{\d\tau}\right)^2+r^2\left(\frac{\d\varphi}{\d\tau}\right)^2\right]\\
&=&-\frac{1}{2},
\end{eqnarray}
where we have imposed the normalization condition of four-velocity in the last step. In terms of the conserved specific energy and specific angular momentum,
\begin{equation}\label{el}
\varepsilon=-g_{tt}\frac{\d t}{\d\tau}=B\frac{\d t}{\d\tau},~~~~\ell=g_{\varphi\varphi}\frac{\d\varphi}{\d\tau}=r^2\frac{\d\varphi}{\d\tau},
\end{equation}
respectively, Eq. \eqref{ham} can be transformed into the radial equation of motion,
%\cite{Zuluaga:2021vjc,Kurmanov:2024hpn}
%\begin{equation}
%\frac{1}{2}\left[-\frac{\varepsilon^2}{B(r)}+\frac{W(r)^2}{B(r)}\left(\frac{\d r}{\d\tau}\right)^2+\frac{\ell^2}{r^2}\right]=-\frac{1}{2}.
%\end{equation}
\begin{equation}\label{reom}
\frac{1}{2}\varepsilon^2=-\frac{1}{2}g_{tt}g_{rr}\left(\frac{\d r}{\d\tau}\right)^2+V_{\eff}(r)=\frac{1}{2}W^2\left(\frac{\d r}{\d\tau}\right)^2+V_{\eff}(r).
\end{equation}
In this equation, the effective potential induced by the geometry is
\begin{equation}\label{Veff}
V_{\eff}(r)=-\frac{1}{2}g_{tt}\left(1+\frac{\ell^2}{g_{\varphi\varphi}}\right)=\frac{B}{2}\left(1+\frac{\ell^2}{r^2}\right).
\end{equation}

The ISCO is a marginally stable circular orbit. For circular obits, we have $\d r/\d\tau=\d^2r/\d\tau^2=0$. The marginal stability condition is $\d^3r/\d\tau^3=0$. Applied to Eq. \eqref{reom}, they are equivalent to
\begin{equation}\label{VdVddV}
V_{\eff}(r)=\frac{1}{2}\varepsilon^2,~~~~V'_{\eff}(r)=0,~~~~V''_{\eff}(r)=0.
\end{equation}
Since we are interested in the ISCO radius, we eliminate $\ell$ in the latter two equations of \eqref{VdVddV} and obtain
\begin{equation}
\left[\frac{1}{g'_{tt}}\left(\frac{g_{tt}}{g_{\varphi\varphi}}\right)'\right]'=0.
\end{equation}
Therefore, the radius of ISCO around the black hole \eqref{metric} is dictated by
\begin{equation}
BB''-2B'^2+\frac{3}{r}BB'=0.
\end{equation}
Insertion of Eq. \eqref{B} and further simplifications yield
\begin{comment}
\begin{eqnarray}
\nonumber q^2 \left(q^4+4 r^2\right) \left(3 q^6 r^4 v-r^8 \left(q^4+2 r^2\right)+\sqrt{q^4+4 r^2} \left(12 q^6 v^2+q^2 r^8+v \left(3 q^2 r^6-3 q^4 r^4\right)\right)\right)&&\\
+2 q^4 r^8 v+3 r^{12} \sqrt{q^4+4 r^2}-r^8 \left(q^2 r^4+r^6\right)
&=&0.
\end{eqnarray}
\begin{eqnarray}
\nonumber&&q^4\left[12q^4v^2+r^8+3r^4v\left(r^2-q^2\right)\right]\left(q^4+4r^2\right)^{3/2}\\
\nonumber&&+q^2r^4\left[3q^6v-r^4\left(q^4+2r^2\right)\right]\left(q^4+4r^2\right)\\
&&+3r^{12}\sqrt{q^4+4 r^2}+r^8\left[2q^4v-r^4\left(q^2+r^2\right)\right]=0.
\end{eqnarray}
\begin{eqnarray}
\nonumber&&Q^4\left[12Q^4(\alpha-\beta)^2+r^8+3r^4(\alpha-\beta)\left(r^2-Q^2\right)\right]\left(Q^4+4r^2\right)^{3/2}\\
\nonumber&&+Q^2r^4\left[3Q^6(\alpha-\beta)-r^4\left(Q^4+2r^2\right)\right]\left(Q^4+4r^2\right)\\
&&+3r^{12}\sqrt{Q^4+4 r^2}+r^8\left[2Q^4(\alpha-\beta)-r^4\left(Q^2+r^2\right)\right]=0,
\end{eqnarray}
\end{comment}
\begin{eqnarray}\label{isco}
\nonumber&&3Q^4\left[4Q^4(\alpha-\beta)^2+r^4(\alpha-\beta)\left(r^2-Q^2\right)\right]\left(Q^4+4r^2\right)^{3/2}\\
\nonumber&&+Q^2r^4\left[3Q^6(\alpha-\beta)-r^4\left(Q^4+2r^2\right)\right]\left(Q^4+4r^2\right)\\
\nonumber&&+r^8\left(Q^8+4Q^4r^2+3r^4\right)\sqrt{Q^4+4 r^2}\\
&&+r^8\left[2Q^4(\alpha-\beta)-r^4\left(Q^2+r^2\right)\right]=0,
\end{eqnarray}
where we have set $M=1$.

\begin{figure}[htbp]
    \centering
    \includegraphics[width=0.45\textwidth]{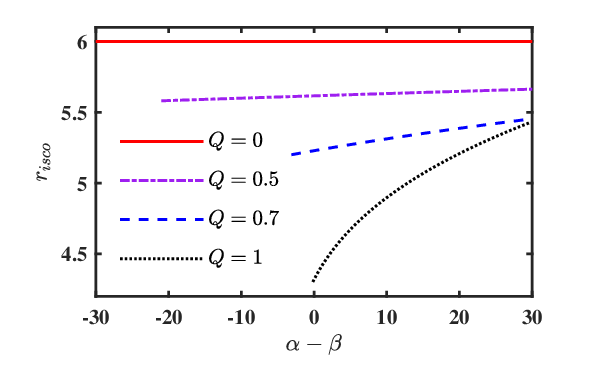}\\
    \includegraphics[width=0.45\textwidth]{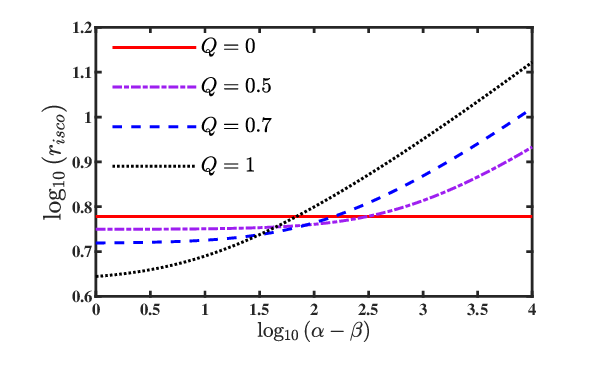}
    \caption{Dependence of the ISCO radius on model parameters $\alpha-\beta$ for different values of magnetic charge $Q$. In the top panel, both axes are scaled linearly, and the start points of curves are dictated by subextremality condition \eqref{Qext}. In the bottom panel, restricted to the region $\alpha-\beta>0$, both axes are in logarithmic scales.
    }\label{fig:isco}
\end{figure}

We failed in finding the analytical solution of this equation and then turned to solve it numerically. The results are shown in Fig. \ref{fig:isco}, from which one can see that a black hole with larger magnetic charge tends to have a smaller ISCO when $\alpha-\beta$ is negative or small, but tends to have a larger ISCO for very large values of $\alpha-\beta$. Especially, when the charge $Q$ vanishes, the ISCO radius reproduces the constant value $\risco=6M$ for Schwarzschild black holes. Assigning a nonzero value to the charge, we find the ISCO radius increases with $\alpha-\beta$ and goes up like $\risco\rightarrow[24Q^4(\alpha-\beta)]^{1/5}$ at $\alpha-\beta\gg|Q|$. The readers can confirm this result by checking the cancellation of leading terms in Eq. \eqref{isco}.

\subsection{General circular orbits}\label{subsect-sco}
Outside the ISCO, all circular orbits are stable, unless there exists an outermost stable circular orbit. For general circular orbits, only the former two equations of \eqref{VdVddV} are required, and we can use them to determine the specific energy $\varepsilon$ and the specific angular momentum $\ell$ of a test particle in a circular orbit of radius $r$,
\begin{eqnarray}
\label{varepsilon}\varepsilon&=&\frac{-g_{tt}}{\sqrt{-g_{tt}+g_{\varphi\varphi}\frac{g'_{tt}}{g'_{\varphi\varphi}}}}=\frac{B}{\sqrt{B-\frac{1}{2}rB'}},\\
\label{ell}\ell&=&\frac{g_{\varphi\varphi}\sqrt{-\frac{g'_{tt}}{g'_{\varphi\varphi}}}}{\sqrt{-g_{tt}+g_{\varphi\varphi}\frac{g'_{tt}}{g'_{\varphi\varphi}}}}=\frac{r\sqrt{\frac{1}{2}rB'}}{\sqrt{B-\frac{1}{2}rB'}}.
\end{eqnarray}
The angular velocity of the particle in this orbit is
\begin{equation}\label{Omega}
\Omega=\frac{\d \varphi}{\d t}=\frac{-g_{tt}\ell}{g_{\varphi\varphi}\varepsilon}=\sqrt{-\frac{g'_{tt}}{g'_{\varphi\varphi}}}=\sqrt{\frac{B'}{2r}},
\end{equation}
where we have made use of Eqs. \eqref{el}, \eqref{varepsilon}, \eqref{ell}.

\begin{figure}[htbp]
    \centering
    \includegraphics[width=0.45\textwidth]{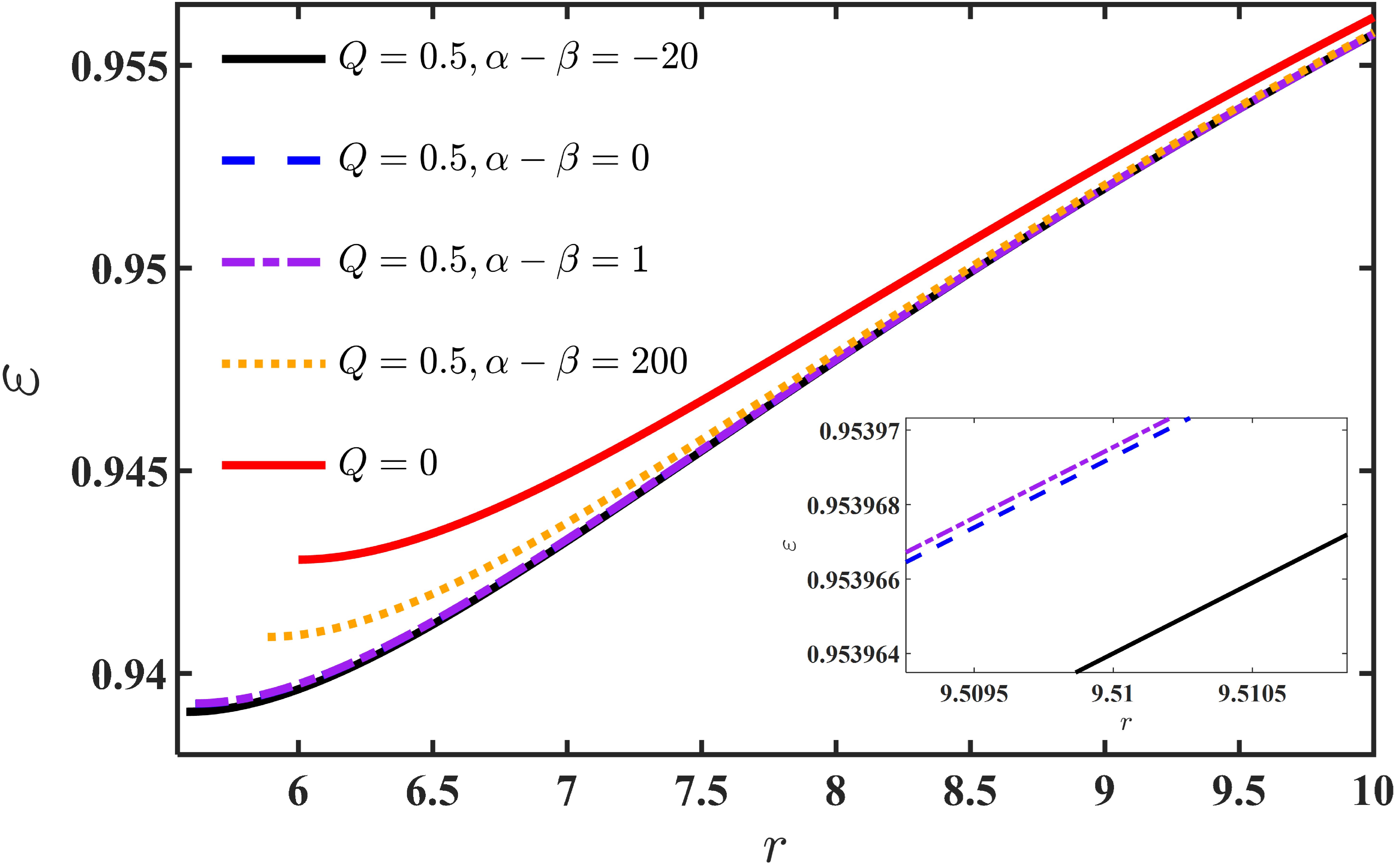}\\
    \includegraphics[width=0.45\textwidth]{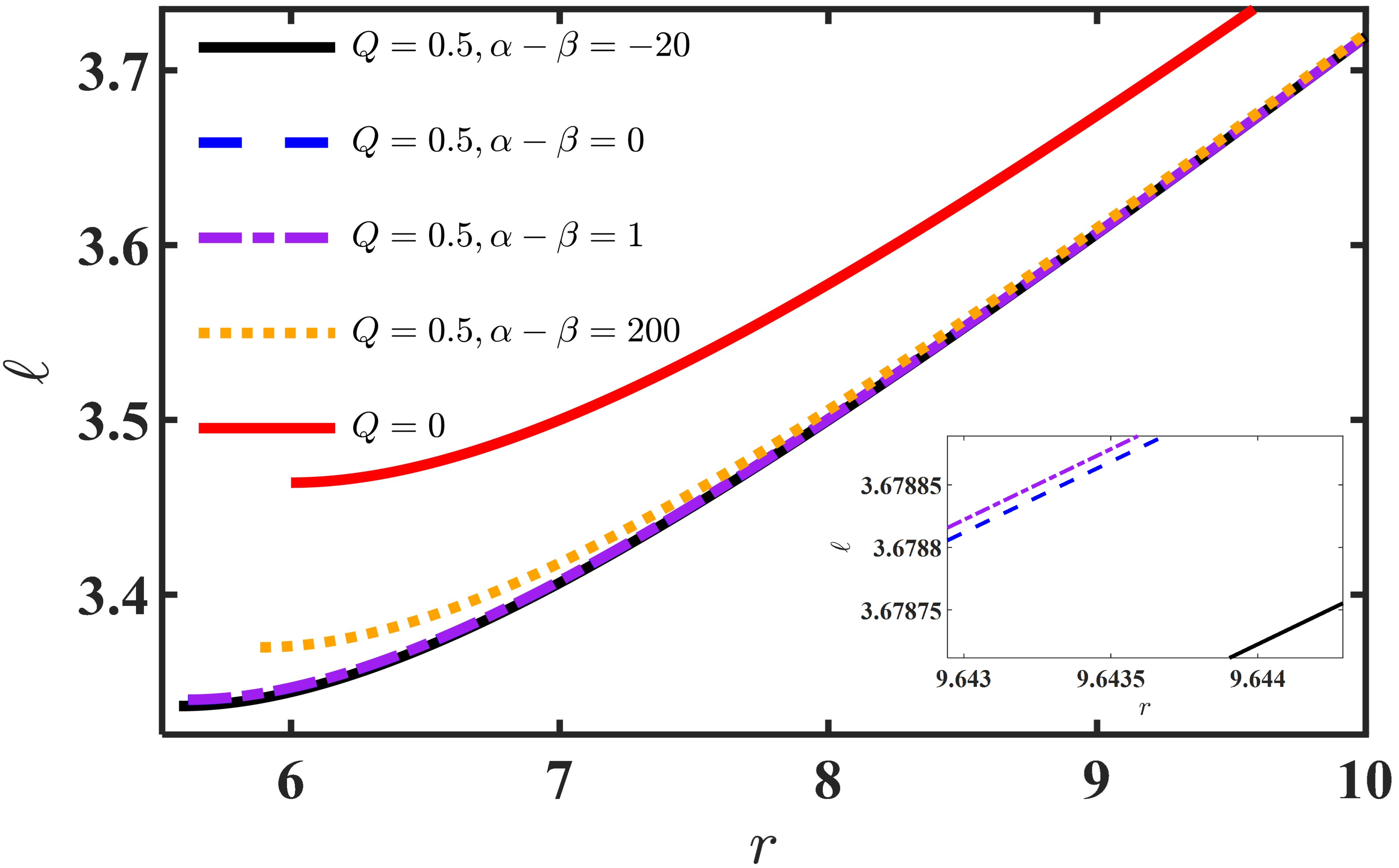}\\
    \includegraphics[width=0.45\textwidth]{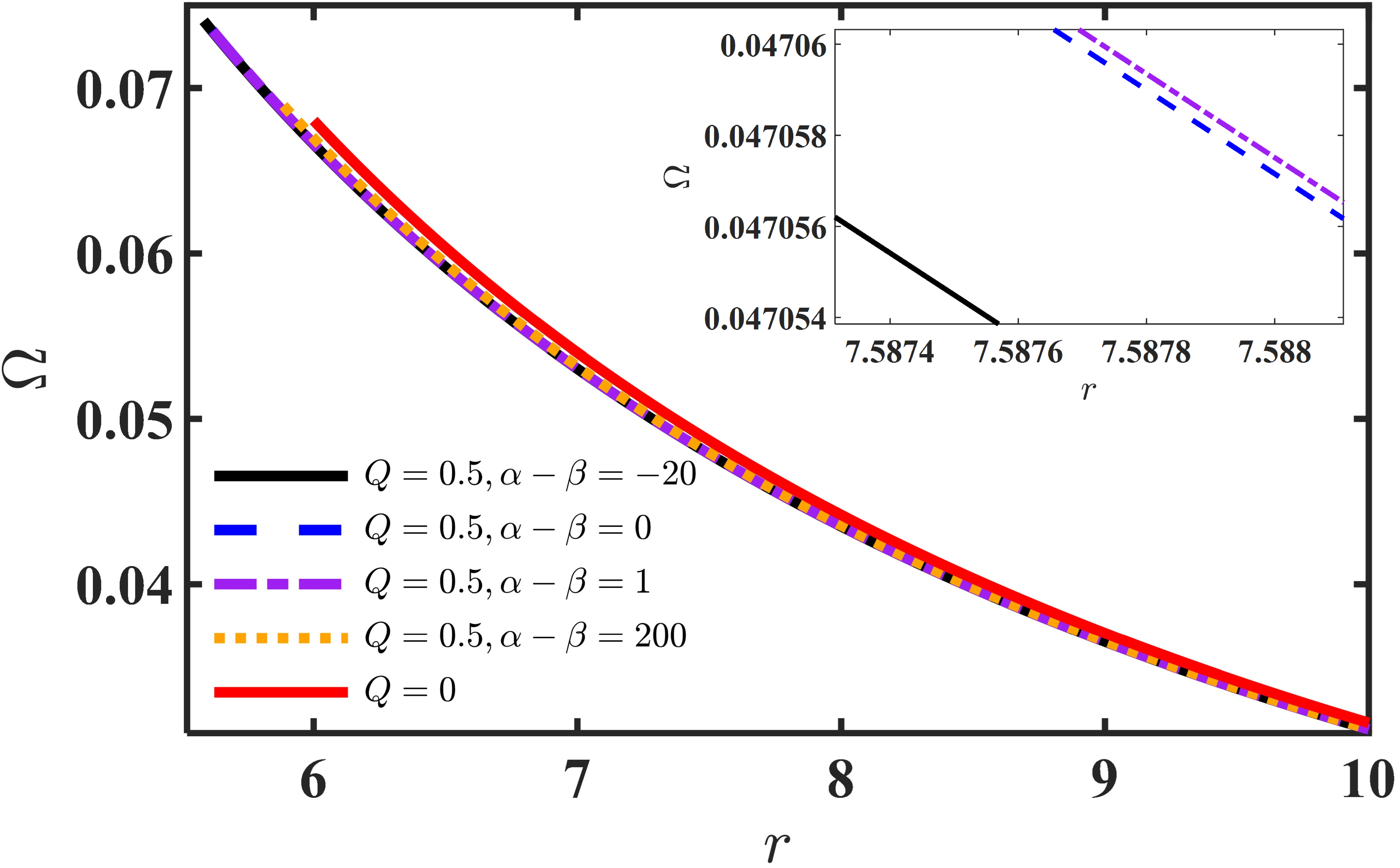}
    \caption{Radial dependence of specific energy (top panel), specific angular momentum (middle panel) and angular velocity (bottom panel) of the particle for various combinations of charge $Q$ and difference $\alpha-\beta$. Each curve starts at the left from $r=\risco$, whose value is in accordance with Eq. \eqref{isco} and Fig. \ref{fig:isco}.
    }\label{fig:spec}
    %}\label{fig:varepsilon}
\end{figure}

Akin to Kepler's third law, in the steady-state thin disk model, the specific energy $\varepsilon$, the specific angular momentum $\ell$ and the angular velocity $\Omega$ of the particle are not arbitrary but dictated simply by the radius of its orbit via Eqs. \eqref{varepsilon}, \eqref{ell} and \eqref{Omega}. Setting $M=1$, we depict them in top, middle and bottom panels of Fig. \ref{fig:spec} in order. The influences of $Q$ and $\alpha-\beta$ on them are more obvious at small radii. From typical examples in the figure, we can see the curves for $\alpha-\beta>0$ lie between the Schwarzschild limit (red solid curves) and the GMGHS limit (blue dashed curves), whereas the curves for $\alpha-\beta<0$ fall slightly below the GMGHS limit in this region. Outside the ISCO, both specific energy and specific angular momentum increase with radius $r$, whereas the angular velocity decreases with radius in all cases.
%\begin{figure}[htbp]
    %\centering
    %\includegraphics[width=0.45\textwidth]{L_q.png}\includegraphics[width=0.45\textwidth]{L_v.png}
    %\caption{specific angular momentum of the particle.
    %}\label{fig:ell}
%\end{figure}
%\begin{figure}[htbp]
    %\centering
    %\includegraphics[width=0.45\textwidth]{O_q.png}\includegraphics[width=0.45\textwidth]{O_v.png}
    %\caption{angular velocity of the particle.
    %}\label{fig:Omega}
%\end{figure}

\begin{figure}[htbp]
    \centering
    \includegraphics[width=0.45\textwidth]{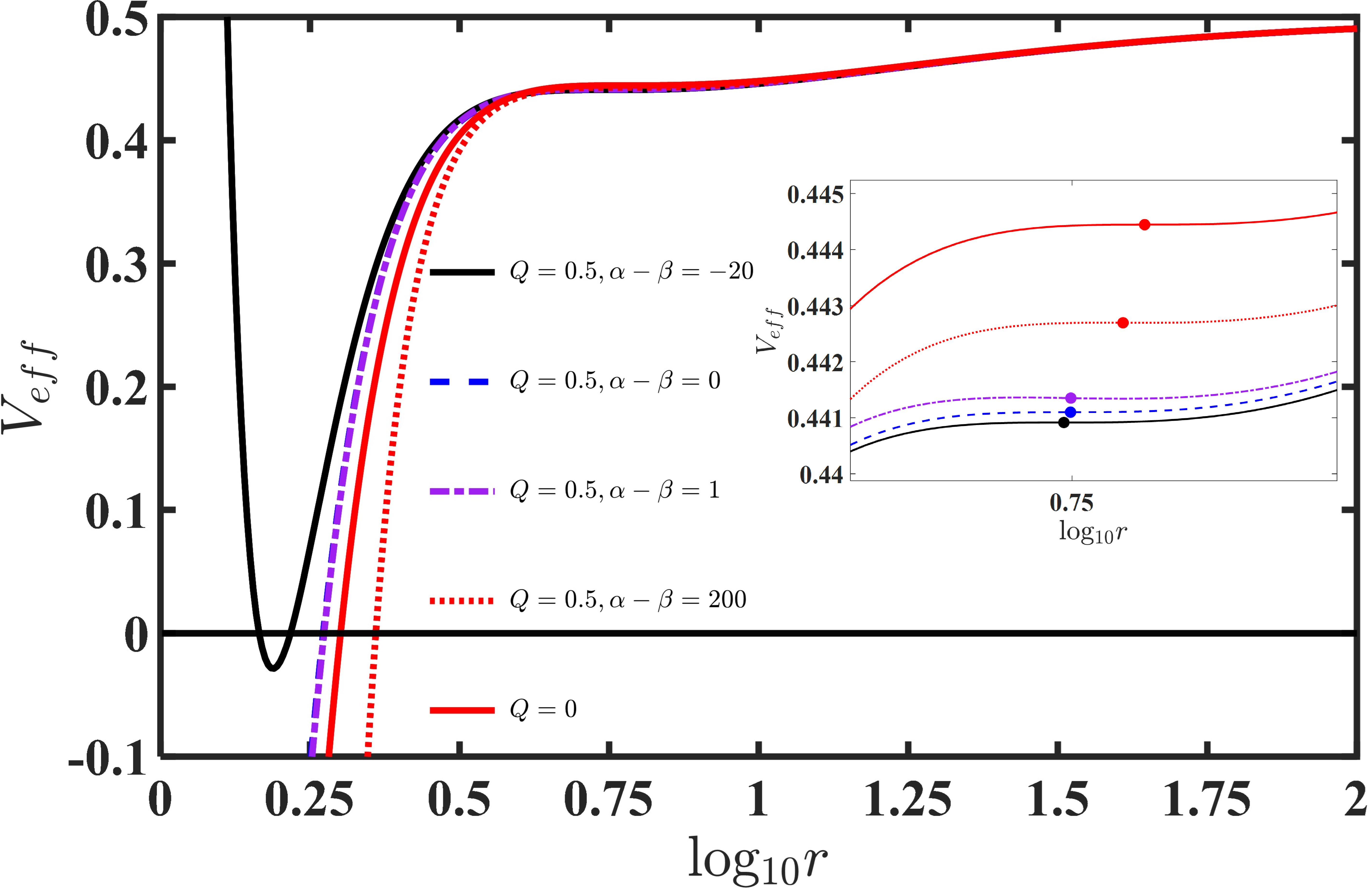}
    \caption{Effective potential $V_{\eff}(r,\ell_{\isco})$ versus $r$, where $\ell$ is evaluated at the ISCO for each combination of $Q$ and $\alpha-\beta$. The locations of ISCO are marked by thick dots.
    }\label{fig:Veff}
\end{figure}

In particular, for each combination of $Q$ and $\alpha-\beta$, we can determine the value of $\risco$ with Eq. \eqref{isco}, and then $\ell_{\isco}$ with Eq. \eqref{ell}. Inserting such values of $Q$, $\alpha-\beta$ and $\ell_{\isco}$ into Eq. \eqref{Veff}, we plot the curve of effective potential $V_{\eff}(r,Q,\alpha-\beta,\ell_{\isco})$ in Fig. \ref{fig:Veff}. Corresponding to five cases listed in Sec. \ref{sect-ec}, there are five curves in Fig. \ref{fig:Veff}. The locations of $r=\risco$ are highlighted by thick dots. Indeed, they are infection points of these curves.

\begin{figure}[htbp]
    \centering
    \includegraphics[width=0.45\textwidth]{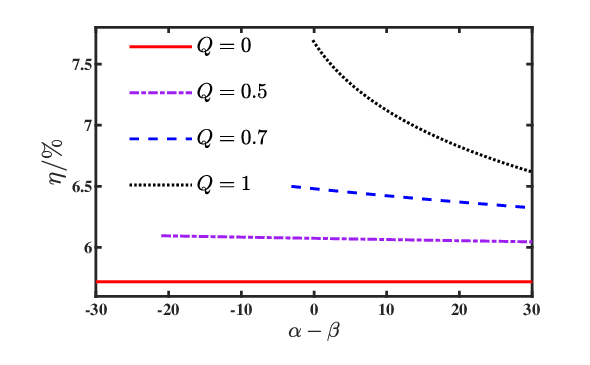}\\
    \includegraphics[width=0.45\textwidth]{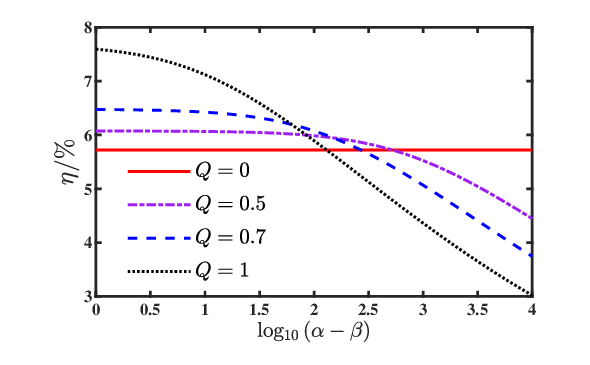}
    \caption{Conversion efficiency of accreting mass into radiation. For four typical values of magnetic charge $Q$, its dependence on parameter difference $\alpha-\beta$ is shown in linear (top panel) and semi-logarithmic (bottom panel) graphs.
    }\label{fig:eta}
\end{figure}

The value of $\varepsilon$ at the ISCO is also of importance. Under the assumption that all emitted photons can escape to infinity, it can be used to estimate the efficiency $\eta$ that the black hole converts the accreting mass into radiation \cite{Thorne:1974ve},
%.Provided that all the emitted photons can escape to infinity, the radiation energy is obtained from the energy loss by a test particle moving from infinity at rest to the inner edge of the disk.
\begin{equation}\label{eta}
\eta=\frac{\mathcal{L}_{\bol}}{\dot{M}c^2}\simeq1-\varepsilon_{\isco},
\end{equation}
where $\mathcal{L}_{\bol}$ is the bolometric luminosity of the accretion disk, $\dot{M}$ is the time-averaged rate of mass accretion, and $\varepsilon_{\isco}$ is the specific energy evaluated at the ISCO. The numerical results are shown in Fig. \ref{fig:eta}. When the value of $\alpha-\beta$ is small (or larger), a black hole with smaller magnetic charge has a higher (or lower) conversion efficiency. Especially, when the charge $Q$ vanishes, the efficiency goes back to the constant $\eta=1-2\sqrt{2}/3$ for Schwarzschild black holes. Assigning a nonzero value to the charge, we find the conversion efficiency decreases with $\alpha-\beta$ and gradually vanishes like $\eta\rightarrow(5/12)[24Q^4(\alpha-\beta)]^{-1/5}$ at $\alpha-\beta\gg|Q|$.

\section{Radiative properties of accretion disks}\label{sect-rad}
For an optically thick accretion disk with negligible radial heat transport, each local patch of the disk is in thermodynamical equilibrium, then the radiation emitted from the disk can be taken as a black-body radiation with a position-dependent temperature. In the current section, we will make use of such a local black-body approximation and conservation equations to study the radiative properties of the geometrically thin, optically thick accretion disk around the black hole \eqref{metric}.

According to the rest mass conservation equation, the time-averaged rate of mass accretion $\dot{M}$ of the black hole is related to the surface density of disk $\Sigma(r)$ via \cite{Feng:2024iqj}
\begin{equation}
\dot{M}=\frac{\d M}{\d t}=-2\pi\sqrt{-g_{tt}g_{rr}g_{\varphi\varphi}}\Sigma(r)\frac{\d r}{\d\tau}.
\end{equation}
%Combined with Eq. \eqref{reom}, it gives the expression of surface density
%\begin{equation}
%\Sigma=-\frac{\dot{M}}{2\pi\sqrt{g_{\varphi\varphi}}}\left(\varepsilon^2-1-2V_{\eff}\right)^{-1/2}=-\frac{\dot{M}}{2\pi r}\left(\varepsilon^2-1-2V_{\eff}\right)^{-1/2},
%\end{equation}
Apparently, if all materials in the disk are moving in perfect circular orbit, i.e. $\d r/\d\tau=0$, then the mass accretion rate will be zero. In a disk with a nonzero accretion rate, particles lose their energy and angular momentum gradually and fall into the black hole eventually. In this process, the energy loss of particles is converted to the radiation energy. As such, the energy flux emitted from the disk can be computed from the equations of energy conservation and angular momentum conservation \cite{Feng:2024iqj}, and the result is \cite{Karimov:2018whx,Alloqulov:2024zln,Feng:2024iqj}
\begin{eqnarray}\label{flux}
\nonumber\mathcal{F}(r)&=&-\frac{\dot{M}}{4\pi\sqrt{-g_{tt}g_{rr}g_{\varphi\varphi}}}\frac{\Omega_{,r}}{\left(\varepsilon-\Omega\ell\right)^2}\int_{\risco}^r\left(\varepsilon-\Omega\ell\right)\ell_{,r}\d r\\
\nonumber&=&-\frac{\dot{M}\left(rB''-B'\right)}{4\pi r^2W\sqrt{2rB'}\left(2B-rB'\right)}\\
&&\times\int_{\risco}^r\frac{\sqrt{r}\left(rBB''-2rB'^2+3BB'\right)}{\sqrt{2B'}\left(2B-rB'\right)}\d r.
\end{eqnarray}
%In the same spirit of Eq. \eqref{eta}, the total luminosity emitted between radius $r$ and infinity can be taken to be proportional to the total energy drop from infinity to $r$. Therefore,
Furthermore, the differential of luminosity $\mathcal{L}_{\infty}$ that reaches spatial infinity from radius $r$ is related to the flux by \cite{Novikov:1973kta,Page:1974he,Thorne:1974ve}
\begin{eqnarray}\label{dL}
\nonumber\frac{\d\mathcal{L}_{\infty}}{d\ln r}&=&4\pi r\sqrt{-g_{tt}g_{rr}g_{\varphi\varphi}}\varepsilon\mathcal{F}(r)\\
\nonumber&=&-\frac{\dot{M}B\left(rB''-B'\right)}{\sqrt{rB'}\left(2B-rB'\right)^{3/2}}\\
&&\times\int_{\risco}^r\frac{\sqrt{r}\left(rBB''-2rB'^2+3BB'\right)}{\sqrt{2B'}\left(2B-rB'\right)}\d r.
\end{eqnarray}

\begin{figure}[htbp]
    \centering
    \includegraphics[width=0.45\textwidth]{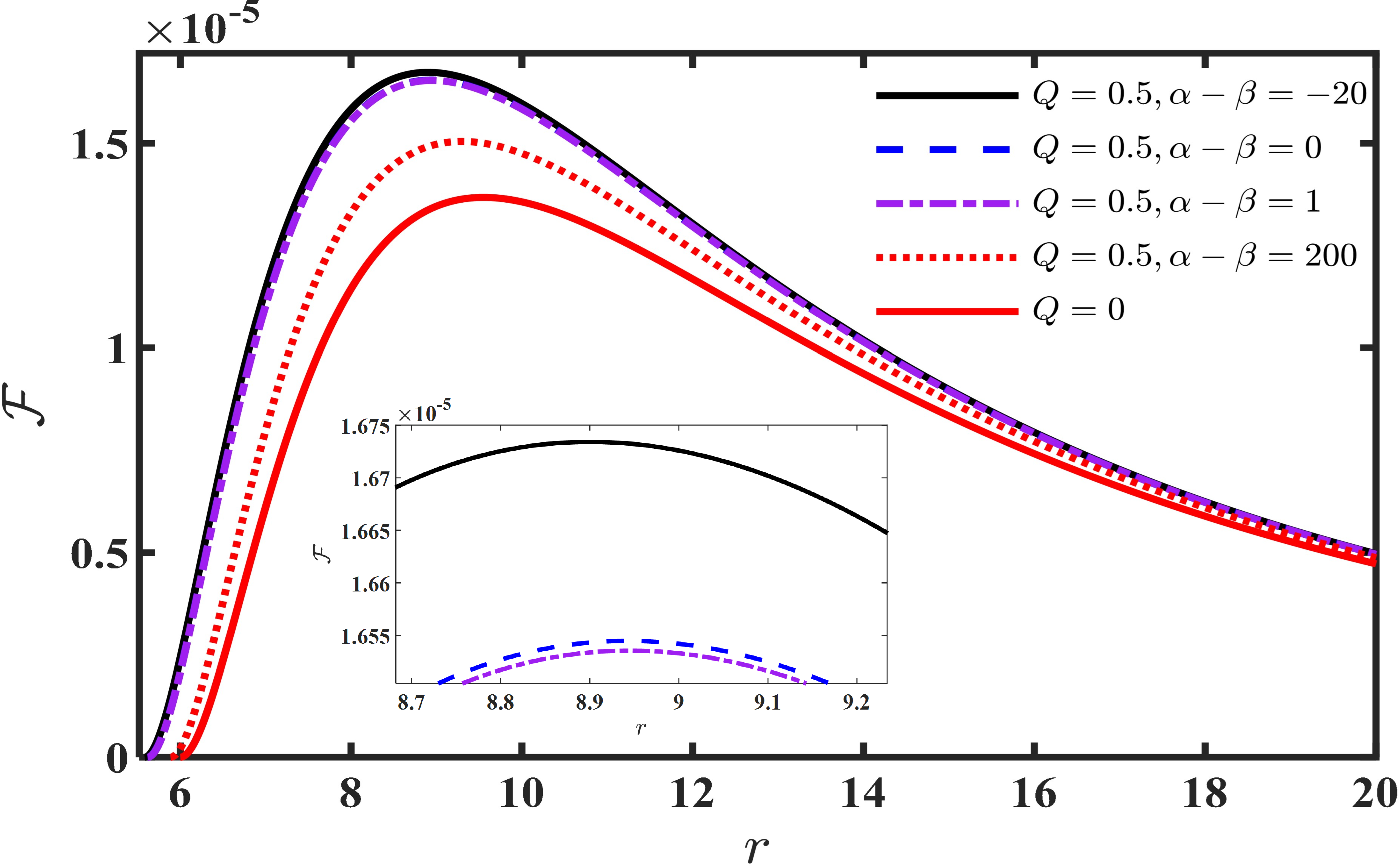}\\
    \includegraphics[width=0.45\textwidth]{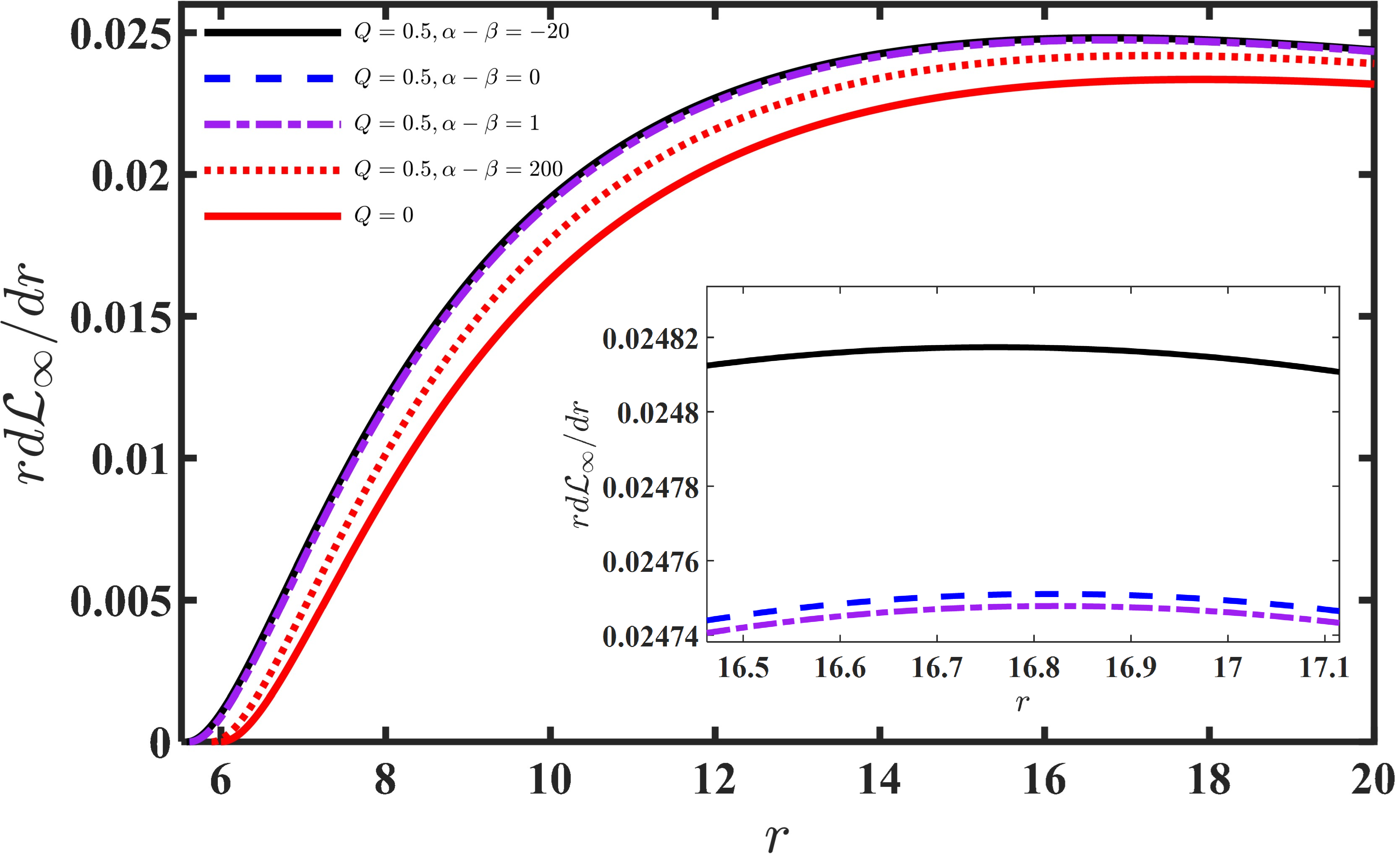}
    \caption{Energy flux (top panel) and differential luminosity (bottom panel) per unit accretion rate.
    }\label{fig:flux}
\end{figure}
Divided by $\dot{M}$, the energy flux \eqref{flux} and the differential luminosity \eqref{dL} have been numerically simulated in Fig. \ref{fig:flux}. It turns out that both of them are enhanced by the magnetic charge at small radii, more significantly in models with smaller values of $\alpha-\beta$. This is mainly attributed to the shift of the ISCO radius, the lower limit of integration in Eqs. \eqref{flux} and \eqref{dL}. Along this line, differences between curves in Fig. \ref{fig:flux} can be well explained by behaviors of $\risco$ in Fig. \ref{fig:isco}. As we have emphasized at the end of Sec. \ref{sect-ec}, our results are unchanged by reversing the sign of $Q$, for instance, by replacing $Q=0.5$ with $Q=-0.5$.

As a simple model, one can apply the Stefan-Boltzmann law $\mathcal{F}(r)=\sigma T_{\eff}^4(r)$ to radiation from the disk in local thermodynamical equilibrium \cite{Page:1974he}. Here $T_{\eff}(r)$ is the local effective temperature on the disk, while $\sigma=2\pi^5k^4/(15c^2h^3)$ is the Stefan-Boltzmann constant. If we capture the gravitational redshift and the Doppler effect with a redshift factor $z$ as
\begin{equation}\label{zsh}
1+z=\frac{1+\Omega r\sin\varphi\sin i}{\sqrt{-g_{tt}-g_{\varphi\varphi}\Omega^2}}=\frac{1+\sqrt{\frac{1}{2}rB'}\sin\varphi\sin i}{\sqrt{B-\frac{1}{2}rB'}}
\end{equation}
and neglect light bending, then the temperature detected by a distant observer can be expressed as \cite{Karimov:2018whx,Bhattacharyya:2000kt}
\begin{equation}\label{Tinf}
T_{\infty}(r)= \frac{T_{\eff}(r)}{1+z}=\frac{\sigma^{-1/4}\mathcal{F}^{1/4}(r)\sqrt{B-\frac{1}{2}rB'}}{1+\sqrt{\frac{1}{2}rB'}\sin\varphi\sin i},
\end{equation}
and the emitted frequency $\nu_e$ will be redshifted to $\nu=\nu_e/(1+z)$. In the above, $i$ is the inclination angle of the disk, and $\varphi$ is the azimuthal coordinate in line element \eqref{metric}.

\begin{figure*}[htbp]
    \centering
    \includegraphics[width=0.35\textwidth]{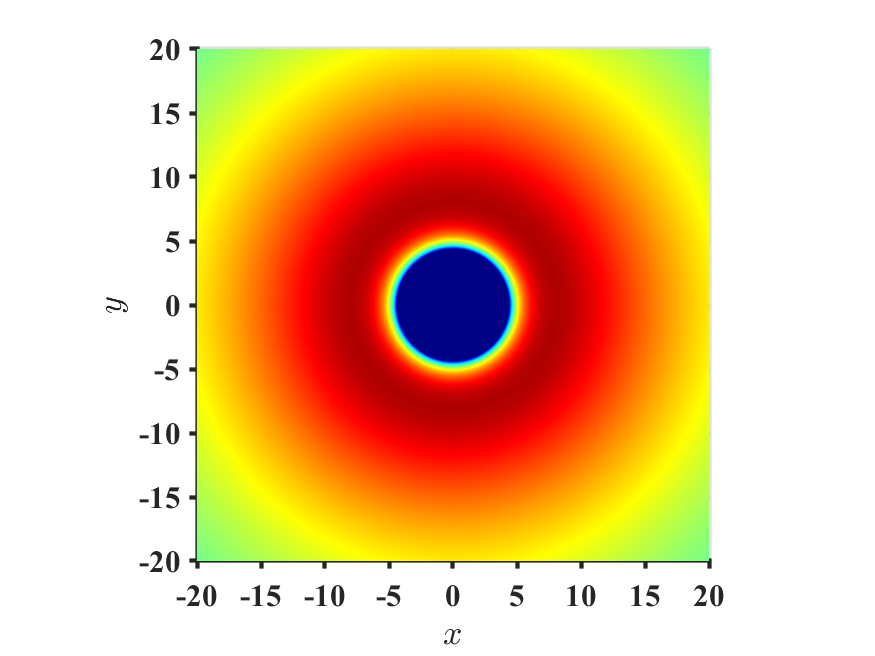}\hspace{-0.06\textwidth}\includegraphics[width=0.35\textwidth]{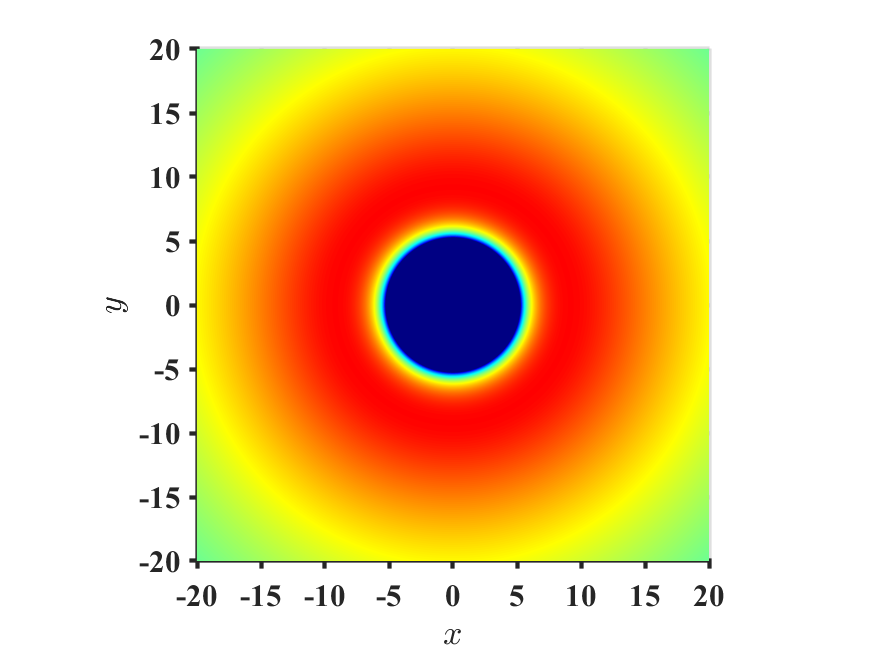}\hspace{-0.05\textwidth}\includegraphics[width=0.35\textwidth]{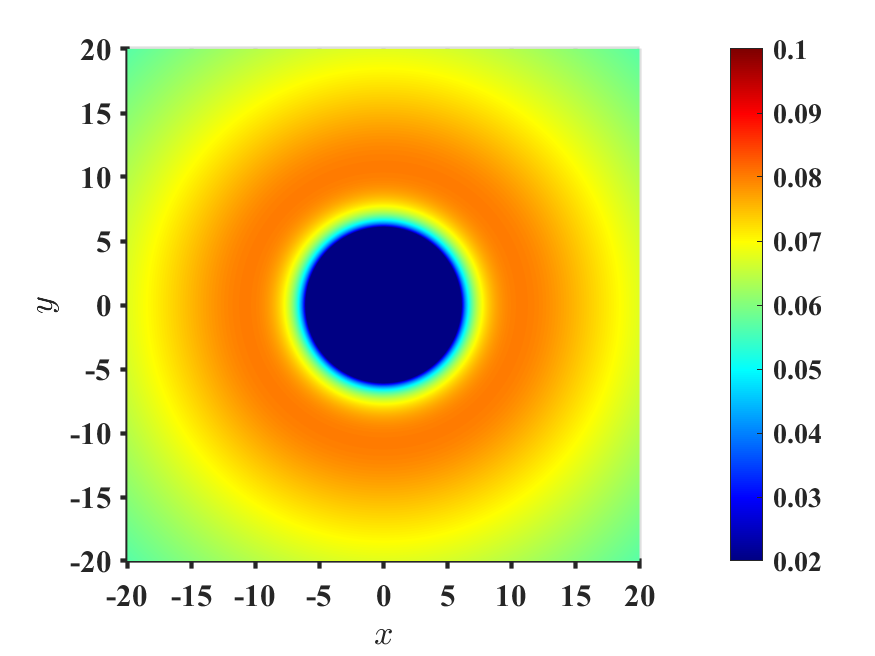}\\
    \includegraphics[width=0.35\textwidth]{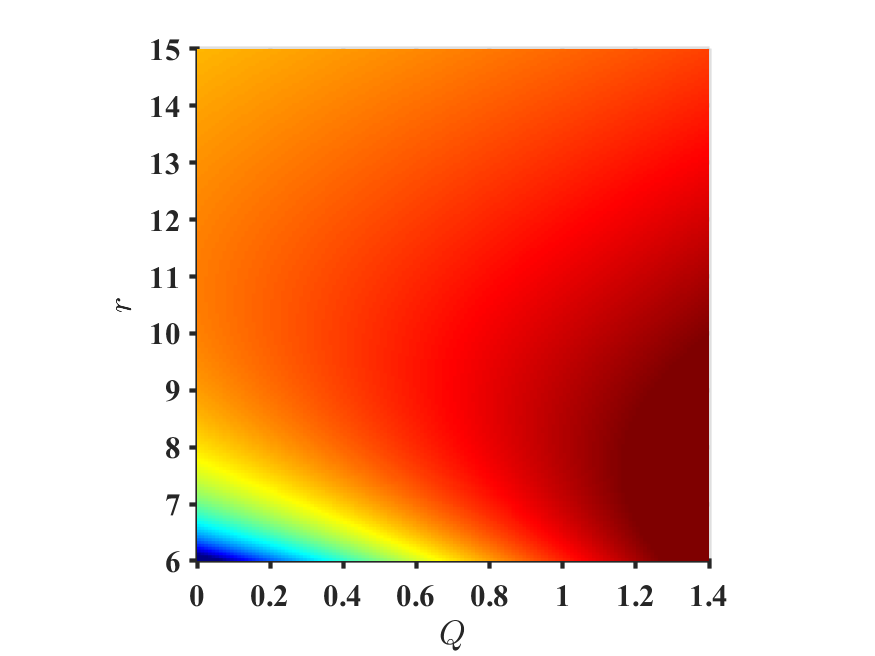}\hspace{-0.06\textwidth}\includegraphics[width=0.35\textwidth]{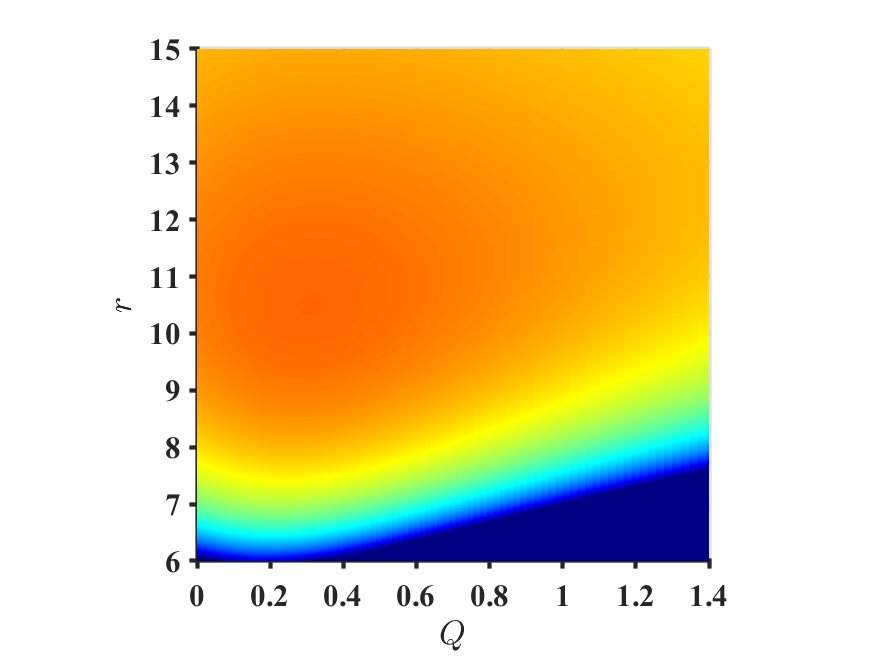}\hspace{-0.05\textwidth}\includegraphics[width=0.35\textwidth]{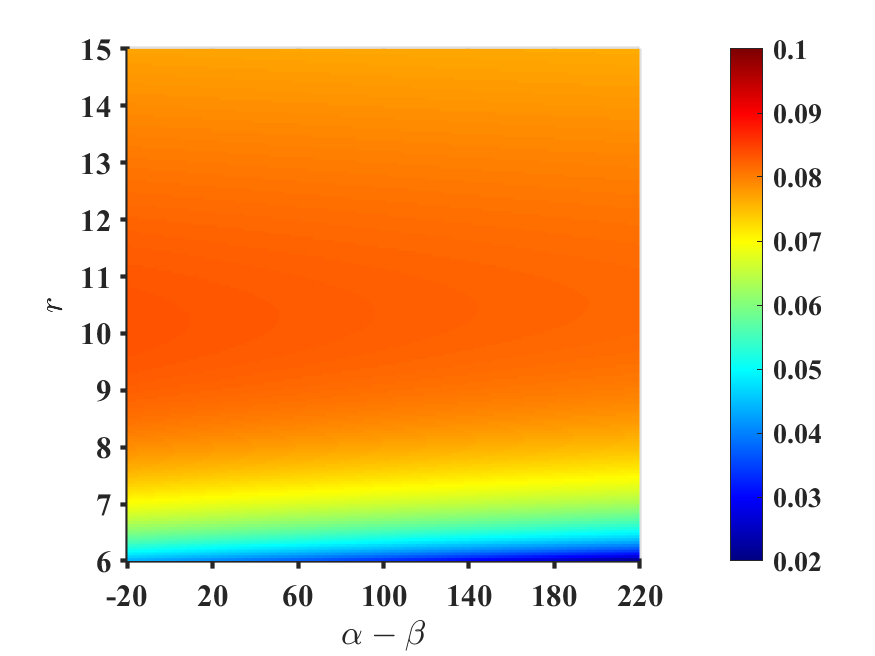}
    \caption{Contour maps of temperature $T_{\infty}(r)$ for a disk with an inclination angle $i=0^{\circ}$. In upper panels, the axes are defined as $x=r\cos\varphi$, $y=r\sin\varphi$, and the parameters are set from left to right as: $Q=1$, $\alpha-\beta=0$; $Q=1$, $\alpha-\beta=20$; $Q=0$. From lower left to lower right panels, the parameters are fixed as: $\alpha-\beta=0$; $\alpha-\beta=200$; $Q=0.5$.
    %the Schwarzschild black hole, a $Q=0.5$ GMGHS black hole,
    }\label{fig:T}
\end{figure*}
For a distant observer along the disk axis $i=0^{\circ}$, some contour maps of temperature \eqref{Tinf} are presented in Fig. \ref{fig:T}. In this case, the temperature is independent of azimuthal angle, but it still depends on three quantities: $r$, $Q$, $\alpha-\beta$. One or two of the quantities should be fixed to plot a map. In the upper row of Fig. \ref{fig:T}, we present temperature maps on the disk plane for three value combinations of $Q$ and $\alpha-\beta$. In the lower row, we fix either $Q$ or $\alpha-\beta$ to illustrate the dependence of temperature $T_{\infty}(r)$ on $\alpha-\beta$ or $Q$. As can be seen from this figure, the magnetically charged black hole in string theory with an Euler-Heisenberg correction has a higher temperature than a Schwarzschild black hole of the same mass, and, for the same value of $Q$, its temperature drops with the increasing value of parameter difference $\alpha-\beta$. These features are understandable, because $T_{\infty}(r)$ is mainly affected by the factor $\mathcal{F}^{1/4}(r)$ in Eq. \eqref{Tinf} and in turn $\mathcal{F}(r)$ is affected mostly by the lower limit of integration in Eq. \eqref{flux}. Moreover, the radius of the dark central area in every map is also determined by the lower limit of integration in Eq. \eqref{flux}.

\begin{figure}[htbp]
    \centering
    \includegraphics[width=0.45\textwidth]{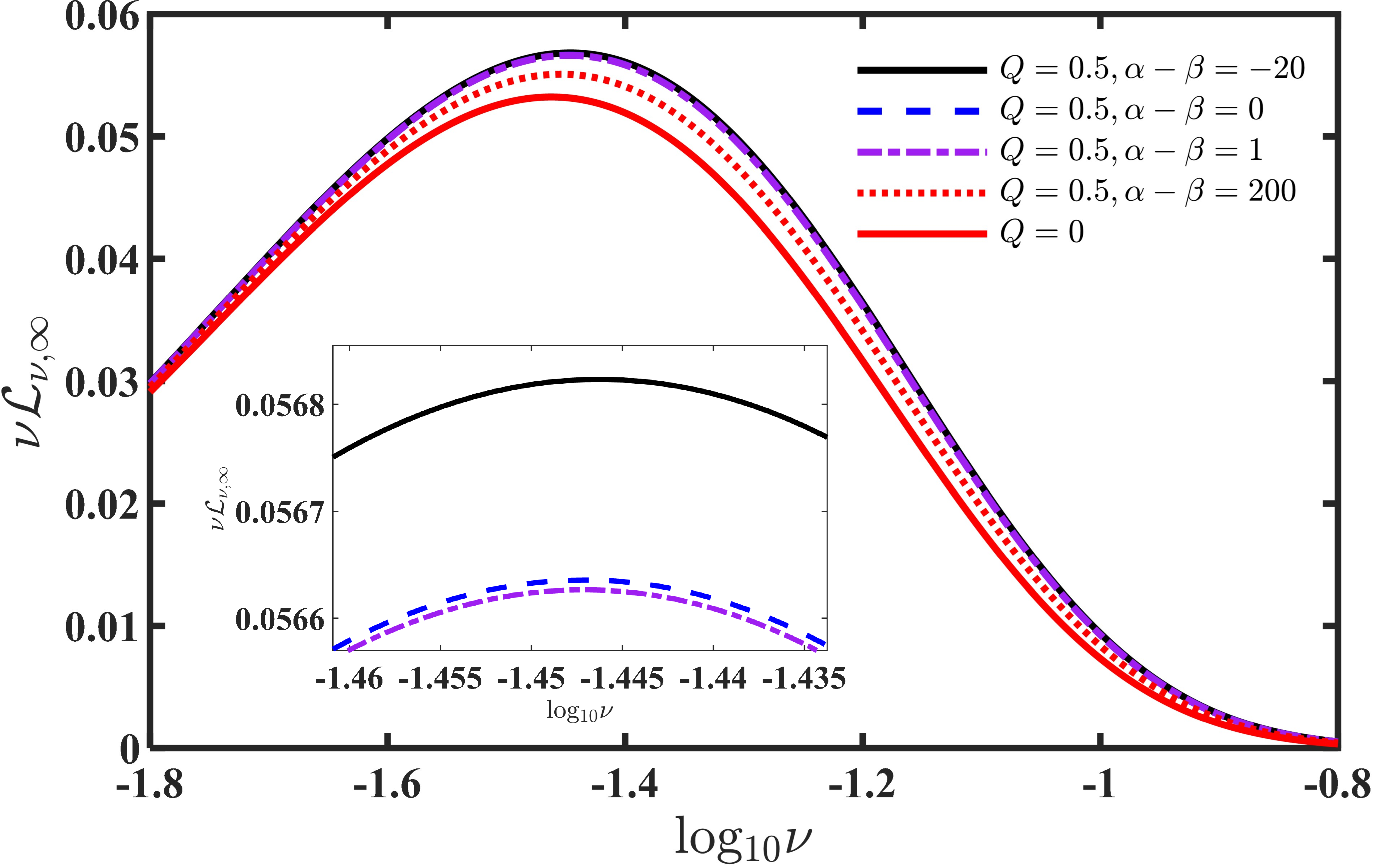}
    \caption{Normalized emission spectrum of the accretion disk with inclination angle $i=0^{\circ}$.\footnote{In the published version of this paper (Phys. Rev. D \textbf{110}, no.10, 103009 (2024)), we have erroneously set $\sigma=\pi^2k^4/(60c^2)$ (which means $\hbar=1$) and $h=1$ in our numerical algorithm. In the present version, we set $\hbar=1$ consistently throughout and replot Fig. \ref{fig:Lnu} accordingly. We are grateful to Shou-Qi Liu for private communications.}
    }\label{fig:Lnu}
\end{figure}

Up to an overall constant factor, the observed specific luminosity $\mathcal{L}_{\nu}$ has a redshifted black-body spectrum \cite{Karimov:2018whx,Bhattacharyya:2000kt,Torres:2002td}
\begin{equation}\label{Lnu}
\mathcal{L}_{\nu,\infty}=4\pi d^2\mathcal{I}_{\nu}=\frac{8\pi h\cos i}{c^2}\int_{\rin}^{\rout}\int_0^{2\pi}\frac{\nu_e^3r\d r\d\varphi}{\exp\left(\frac{h\nu_e}{kT_{\eff}}\right)-1},
\end{equation}
where $\rin=\risco$ and $\rout$ are inner and outer radii of edges of the disk, and $d$ is the distance of the disk center from the observer. Substituting the emitted frequency $\nu_e=\nu(1+z)$ and the local temperature $T_{\eff}=\sigma^{-1/4}\mathcal{F}^{1/4}(r)$ into Eq. \eqref{Lnu} and setting $c=\hbar=k=1$, we draw the spectral curves in Fig. \ref{fig:Lnu} viewed from the angle $i=0^{\circ}$. Although we have set $\rout=100$ in drawing the figure, resetting it to greater values does not lead to perceivable changes. In the low-frequency limit, it is easy to prove that $\exp\left(2\pi\nu_e/T_{\eff}\right)-1\simeq2\pi\nu_e/T_{\eff}$ and thus $\nu\mathcal{L}_{\nu,\infty}\propto \nu^3$ analytically. In the high frequency limit, $1/\left[\exp\left(2\pi\nu_e/T_{\eff}\right)-1\right]\simeq\exp\left(-2\pi\nu_e/T_{\eff}\right)$, the integrand in Eq. \eqref{Lnu} is seriously suppressed. The influences of $Q$ and $\alpha-\beta$ on spectral curves are roughly similar to their influences on energy flux, though Eq. \eqref{Lnu} has more complicated dependence on $\risco$ than Eqs. \eqref{flux} and \eqref{dL}.

\section{Outlook}\label{sect-out}
For magnetically charged black holes in string theory with an Euler-Heisenberg correction \cite{Bakopoulos:2024hah}, we have refined the subextremality condition and energy conditions, and we have studied the Novikov-Thorne model to describe geometrically thin, optically thick accretion disks around them. For null, weak and strong energy conditions to all hold outside event horizons, the necessary and sufficient condition is inequality \eqref{SEC} with $r$ replaced by $\rh$. Besides the specific energy, specific angular momentum and angular velocity of circularly orbiting particles, the energy flux, local temperature, luminosity of the disk and the conversion efficiency of accreting mass into radiation are mainly affected by the ISCO radius, which is taken as the radius of inner edge. When $\alpha-\beta$ is positive but not very large, these quantities and ISCO radius take values between their Schwarzschild ($Q=0$) limit and the GMGHS ($\alpha-\beta=0$) limit. When $\alpha-\beta$ is negative or very large, their values are outside the Schwarzschild limit or the GMGHS limit.

Although the focus of this paper is on theoretical aspects, we look forward to connecting our results to past and future observations, such as the black hole shadow, iron line shape and X-ray reflection spectroscopy \cite{Reynolds:2013qqa,Bambi:2015kza}.

Models of optically thin and geometrically thin disks are popular in studying how the shape of the emission region may affect the observational appearance of black holes \cite{Gralla:2019xty,Qu:2023hsy} in recent years. Compared with the static disk \cite{Gralla:2019xty} and the free-falling disk \cite{Qu:2023hsy}, the circularly rotating disk constructed in this paper is more realistic. It will be interesting to make a comparison of these models by simulating the observational appearance of black hole.

Radiation emitted (reflected) by the accretion disk and a hot electron cloud is exploited in many techniques to test the Kerr black hole hypothesis and to constrain the black hole spin \cite{Reynolds:2013qqa,Bambi:2015kza}. However, as shown by worked non-Kerr examples in the past, the observed spectrum is often degenerate with respect to the spin and other parameters of the black hole \cite{Bambi:2015kza,Bambi:2020cyv}. It will be also interesting to examine the parameter degeneracy in magnetically charged rotating black holes in string theory with an Euler-Heisenberg correction. Unfortunately, the solution discovered in Ref. \cite{Bakopoulos:2024hah} is a nonrotating black hole. To make greater contact with astrophysical observations, it is compelling for us to find the rotating solution in the same theory.

For Kerr black holes, the classic work \cite{Bardeen:1972fi} has provided solutions of specific energy and angular momentum to both corotation and counterrotation accretions. If we find the magnetically charged rotating black holes in string theory with an Euler-Heisenberg correction, then the corotation case and the counterrotation case should be treated with care when studying the properties of their accretion disks. We leave it for future work.

\acknowledgments{This work is supported by the Natural Science Foundation of Shanghai (Grant No. 24ZR1419300).}

%\appendix

\end{document}